\documentclass[aps,prd,twocolumn,superscriptaddress,citeautoscript,nofootinbib,showkeys,preprintnumbers]{revtex4-1}                  
\pdfoutput=1 
\usepackage{amsmath,amssymb,mathrsfs,bm,feynmf,setspace}
\usepackage{amsthm}
\usepackage{graphicx}     
\usepackage[tight]{subfigure}     
\usepackage[absolute,overlay]{textpos}
\usepackage[usenames, dvipsnames]{color} 
\usepackage{braket} 
\usepackage[utf8]{inputenc}
\usepackage{tensor} 
\usepackage[dvipsnames]{xcolor}
\definecolor{cadmiumred}{rgb}{0.89, 0.0, 0.13}
\definecolor{darkblue}{rgb}{0.2, 0, 0.8}
\definecolor{darkgreen}{rgb}{0.2, 0.71, 0}  
\definecolor{LinkGreen}{cmyk}{1,0.05,.95,0.5}
\usepackage[colorlinks=true]{hyperref}  
\hypersetup{
    bookmarks=true,         
    unicode=false,          
    pdftoolbar=true,        
    pdfmenubar=true,        
    pdffitwindow=false,     
    pdfstartview={FitH},    
    pdftitle={The small black hole illusion},    
    pdfnewwindow=true,      
    colorlinks=true,       
    linkcolor=Blue, 
    citecolor=cadmiumred,        
    filecolor=blue,      
    urlcolor=Blue         
} 

\newtheorem*{hypothesis*}{Hypothesis}

\begin{document}

\title{The small black hole illusion}
\author{Pablo A. Cano}
\email{pablo.cano@uam.es}
\affiliation{Instituto de F\'isica Te\'orica UAM/CSIC, Nicol\'as Cabrera 13, C.U. Cantoblanco, E-28049 Madrid, Spain}

\author{Pedro F. Ram\'irez}
\email{ramirez.pedro@mi.infn.it}
\affiliation{INFN, Sezione di Milano, Via Celoria 16, 20133 Milano, Italy}

\author{Alejandro Ruip\'erez}
\email{alejandro.ruiperez@uam.es}
\affiliation{Instituto de F\'isica Te\'orica UAM/CSIC, Nicol\'as Cabrera 13, C.U. Cantoblanco, E-28049 Madrid, Spain}


\preprint{IFT-UAM/CSIC-18-92}
\preprint{IFUM-1064-FT}

\begin{abstract}
Small black holes in string theory are characterized by a classically singular horizon with vanishing Bekenstein-Hawking entropy. It has been argued that higher-curvature corrections resolve the horizon and that the associated Wald entropy is in agreement with the microscopic degeneracy. In this note we study the heterotic two-charge small black hole and question this result, which we claim is caused by a misidentification of the fundamental constituents of the system studied when higher-curvature interactions are present. On the one hand, we show that quadratic curvature corrections do not solve the singular horizon of small black holes. On the other, we argue that the resolution of the heterotic small black hole reported in the literature involves the introduction of solitonic 5-branes, whose asymptotic charge vanishes due to a screening effect induced by the higher-curvature interactions, and a Kaluza-Klein monopole, whose charge remains unscreened.
\end{abstract}   
\maketitle


Consider a fundamental heterotic string carrying winding number $w$ and momentum $n$ charges along a circle $\mathbb S^1$ \cite{Dabholkar:1989jt, Dabholkar:1990yf}, which forms part of the compact space $\mathbb{T}^4 \times \mathbb{S}^1 \times \hat{\mathbb{S}}^1$. For large values of $n$ and $w$, the entropy associated to the degeneracy of these states is

\begin{equation}
\label{eq:microentropy}
S 
\approx 4\pi \sqrt{nw} \, , \qquad n, w>>1\ .
\end{equation}

\noindent
It was soon suggested that this system, among others, could be also represented as a black hole at strong coupling \cite{Susskind:1993ws, Susskind:1994sm}. In this case the black hole would be \emph{small}, because the event horizon scale would be of the order of the string length. 

Working with the heterotic effective action at lowest order in the perturbative expansion, a solution to the equations of motion carrying the same two charges as this configuration was found in \cite{Sen:1994eb}. That solution is characterized by a singular horizon of vanishing area and entropy, so higher-curvature terms in the effective expansion cannot be ignored in the near-horizon region. This just means that the effective classical description fails to give a good approximation of the system. In a seminal article \cite{Sen:1995in} it was then conjectured that the higher-curvature corrections might somehow render the horizon regular, and that the Wald entropy of such black hole would match the microscopic value \eqref{eq:microentropy}.

Always assuming the existence of a regular horizon and making use of the associated attractor mechanism, a precise matching of the macroscopic and microscopic computations of the entropy was later reported in \cite{Dabholkar:2004yr} --- see also \cite{Dabholkar:2004dq}. This correlation was widely interpreted as a proof of the resolution of the horizon previously conjectured. However, in order to firmly establish if there is causation, there are certain aspects of these studies that need clarification.

In the first place, the techniques employed in those articles only allow for the study of the near-horizon regime, but the analytic construction of a full black hole solution interpolating between those and asymptotic Minkowski space is missing. According to numerical studies \cite{Sen:2004dp}, the solution exists and its causal structure is identical to that of four-charge regular black holes \cite{Hubeny:2004ji}. Nevertheless, in order to fully understand the system studied an analytic solution is needed, specially if we take into consideration that higher-curvature corrections introduce significant global interactions \cite{Cano:2018qev}. 

In the second place, it is certainly surprising that the resolution of the small black hole described in \cite{Dabholkar:2004yr, Dabholkar:2004dq, Sen:2004dp, Hubeny:2004ji} is achieved with the inclusion of only curvature squared terms. Since the departing system is singular, there is no reason to expect that further higher-derivative corrections can be disregarded for that purpose. 

And in the third place, the resolution of similar singular systems, like a Type II string with winding and momentum charges, has not been observed. The different behavior of small black holes in diverse theories raises a puzzle whose resolution has remained unclear so far. 

In this note we argue that the resolution of the heterotic small black hole via higher-curvature corrections does not actually occur. To do so, we apply the results of \cite{Chimento:2018kop, Cano:2018brq}, where the analytic construction of general four-charge, supersymmetric black holes including curvature squared terms has been performed. A simple argument illustrates that apparently regular four-dimensional, two-charge black holes contain a curvature singularity when embedded in the heterotic theory. 

We claim that the resolution of the horizon previously reported is an illusion; the higher-curvature corrections do not really resolve the singularity of \cite{Sen:1994eb}. Instead, we find a special four-charge black hole whose entropy is precisely given by $4\pi \sqrt{nw}$ and whose asymptotic S5 brane charge vanishes, but that does not actually describe the two-charge system. We call this solution a \emph{fake} small black hole and we argue that the resolution of the horizon observed in the literature corresponds to this system.

Although this might seem to represent a step back in the understanding of small black holes, we actually believe that our result clarifies the situation. It puts every small black hole at the same qualitative level; the system is non-perturbative and cannot be properly described incorporating a subgroup of the higher-curvature corrections. Moreover, we recall that the heterotic small black hole can be resolved in the type IIB frame using the uncorrected effective action \cite{Lunin:2001jy, Lunin:2002iz}, via smooth geometries whose degeneracy agrees with \eqref{eq:microentropy}, an observation that gave rise to the fuzzball proposal \cite{Mathur:2005zp, Skenderis:2008qn}.

\textbf{The zeroth-order solution:}
Let us start revisiting the small black hole singular solution. We first review four-charge regular black holes and then particularize to the two-charge case. We work directly in the ten dimensional formulation of low energy heterotic string theory. The bosonic field content consists of the metric $g_{\mu\nu}$, the dilaton $e^\phi$ and the Kalb-Ramond 2-form $B_{\mu\nu}$ with field strength $H_{\mu\nu\rho}$. The action and equations of motion, including all relevant terms at quadratic order in curvature, are briefly described in the supplemental material. In our first approach we neglect these higher-curvature corrections, which will be recovered afterwards. The four-charge solution is,

\begin{eqnarray}
\nonumber
ds^{2}
& = &
\frac{2}{\mathcal{Z}_{-}}du
\left[dv-\tfrac{1}{2}\mathcal{Z}_{+} du\right]
-\mathcal{Z}_{0}d\sigma^{2}_{(4)}
-d\vec{y}^2\, ,
\\
& & \nonumber \\ \nonumber
e^{-2{\phi}}
& = &
g_s^{-2}\frac{\mathcal{Z}_{-}}{\mathcal{Z}_{0}}\, ,
\\
& & \nonumber \\
\label{eq:H}
H 
& = & 
d\mathcal{Z}^{-1}_{-}\wedge du \wedge dv+\star_{(4)}d\mathcal{Z}_{0}
\, ,  
\end{eqnarray}

\noindent
where the Hodge dual in the last equation is associated to the four-dimensional metric $d\sigma^2_{(4)}$, which is a Gibbons-Hawking (GH) space:

\begin{equation}
d\sigma^2_{(4)}=\mathcal{V}^{-1} \left( d z + \chi \right)^2+
\mathcal{V} d\vec{x}^2_{(3)}
\, , \qquad
d \mathcal{V}=\star_{(3)} d \chi \, .
\label{eq:GH}
\end{equation}

\noindent
Six of the coordinates are compact: the coordinates $y^i$ parametrize a four-torus $\mathbb{T}^4$ with no dynamics, while $u$ and $z$ have respective periods $2\pi R_u$ and $2\pi R_z$ and they parametrize two circles that we denote $\mathbb{S}^1$ and $\hat{\mathbb{S}}^1$ .

The functions $\mathcal{Z}_{0,\pm}$ and $\mathcal{V}$ characterize the solution, and are given by 
\begin{eqnarray}
\label{eq:Zs}
\mathcal{Z}_{0,\pm} 
& = & 
1+\frac{q_{0,\pm}}{r}\
\, ,
\quad 
\mathcal{V} 
 = 
1+\frac{q_v}{r}
\end{eqnarray}

\noindent
where $r$ is the radial coordinate of the Euclidean space $dx_{(3)}^2$. Notice that all of these functions are harmonic in this space. This solution represents the superposition of 
\begin{itemize}
\item{a string wrapping the circle $\mathbb{S}^1$ with winding number $w$ and momentum $n$ charges,}
\item{a stack of $N$ solitonic 5-branes (S5) wrapped on $\mathbb{T}^4 \times \mathbb{S}^1$,}
\item{and a Kaluza-Klein monopole (KK) of charge $W$ associated with $\hat{\mathbb{S}}^1$.}
\end{itemize}
The charge parameters $q_i$ are given in terms of the integer numbers $n$, $w$, $N$ and $W$ according to
\begin{eqnarray}
\label{eq:charges}
q_{+}&=&\frac{\alpha'^2 g_{s}^{2}n}{ 2 R_z R_{u}^{2}},\, q_{-}=\frac{\alpha' g_{s}^{2}w}{2 R_z } ,\,
q_0=\frac{\alpha' N}{2 R_z},\,q_v=\frac{W R_z}{2}.
\end{eqnarray}

After compactification in $\mathbb{T}^4\times\mathbb{S}^1\times\hat{\mathbb{S}}^1$, the lower dimensional spacetime metric in the Einstein frame is \cite{Cano:2018brq}
\begin{eqnarray}
\label{eq:metriccomp}
ds_{(4)}^{2}
& = &
\left(\mathcal{Z}_{+} \mathcal{Z}_{-} \mathcal{Z}_{0}\mathcal{V}\right)^{-\frac{1}{2}} dt^2
-\left(\mathcal{Z}_{+} \mathcal{Z}_{-} \mathcal{Z}_{0}\mathcal{V} \right)^{\frac{1}{2}}dx^2_{(3)}\, . 
\end{eqnarray}

\noindent
For non-vanishing charges this geometry represents an extremal black hole whose horizon is placed at $r=0$ and its area is $A\propto \sqrt{nwNW}$. 

The small black hole described in the introduction is that without KK monopole and S5 brane: $N=W=0$. In that case, at $r=0$ there is still a horizon because $g_{tt}$ vanishes. However, its area is zero and, even worse, its curvature diverges. Hence, classically, these solutions have singular horizon and vanishing Bekenstein-Hawking entropy \footnote{This statement holds when any of the four charges vanishes.}. The dilaton $e^{\phi}$ vanishes at the horizon, so loop corrections can be neglected in this region, but the singularity in curvature signals that the tree-level supergravity description of this system is not valid for small values of $r$. When trying to describe the physics near the horizon, one is forced to include the tower of higher-curvature corrections to the heterotic effective action \cite{Bergshoeff:1989de}. For quite some time, it has been believed that their inclusion render the horizon regular and make the value of the Wald entropy of the solution coincide with that of \eqref{eq:microentropy}.
 
Let us discuss how the first set of these corrections, which are quadratic in the curvature, alter relevant aspects of the solutions. These appear at first-order in the $\alpha'$ perturbative expansion of the effective field theory.

\textbf{Two-charge solution at order $\alpha'$:}
A first set of corrections comes from the inclusion of Lorentz and gauge Chern-Simons terms in the Kalb-Ramond field strength, such that its Bianchi identity becomes

\begin{equation}
\label{eq:bianchi}
dH=\frac{\alpha'}{4} \left(
{F}^{A}\wedge{F}^{A}
+
{R}_{(-)}{}^{{a}}{}_{{b}}
\wedge
{R}_{(-)}{}^{{b}}{}_{{a}}
\right)\, .
\end{equation} 

\noindent
In the small black hole literature the gauge fields are set to zero, so we simply take $F^A=0$ from now on. On the other hand, ${R}_{(-)}{}^{{a}}{}_{{b}}$ is the 2-form curvature of the torsionful spin connection, defined as ${\Omega}_{(-)}{}^{{a}}{}_{{b}} ={\omega}^{{a}}{}_{{b}}-\tfrac{1}{2}{H}_{{\mu}}{}^{{a}}{}_{{b}}dx^{{\mu}}$, where ${\omega}^{{a}}{}_{{b}}$ is the spin connection. 

The recursive definition of $H$ introduces a tower of infinite corrections in the perturbative expansion in $\alpha'$ that breaks the supersymmetry of the action, which has to be recovered order by order. At the order we are working, the term $-\frac{\alpha'}{8}{R}_{(-)\,\mu\nu}{}^{{a}}{}_{{b}} {R}_{(-)}{}^{\mu\nu\,{b}}{}_{{a}}$ in the action includes all the corrections quadratic in curvature to the heterotic effective action \footnote{Other corrections unrelated to the supersymmetrization of the Chern-Simons term appear first at fourth-order in curvature.}. Notice that, although we are working at first-order in $\alpha'$ in the action, the form of the Bianchi identity \eqref{eq:bianchi} is exact in this expansion. 

The analysis of the higher-curvature corrections to black hole solutions in the literature 
has been mostly limited to near-horizon geometries. Only very recently, the first-order in $\alpha'$ corrections to the solution (\ref{eq:H})-\eqref{eq:Zs} have been determined \cite{Chimento:2018kop, Cano:2018brq}. First, let us note that the structure of the fields in \eqref{eq:H} is not modified, as that configuration describes general spherically symmetric, asymptotically flat supersymmetric solutions. 
The analysis of the equations of motion reflects that the function $\mathcal{Z}_-$ is not affected by the inclusion of terms quadratic in curvature. It is determined from the Kalb-Ramond 2-form equation, which remains unmodified for this family of solutions. The function $\mathcal{V}$ also remains unmodified, due to supersymmetry constraints. On the other hand, the functions $\mathcal{Z}_{0}$ and $\mathcal{Z}_+$ do get corrected. We have computed the first-order in $\alpha'$ corrections to these functions for any of the zeroth-order solutions presented above. Let us consider the corrections to the zeroth-order solution with $q_v=q_0=0$, which represents the 2-charge system. In that case, we get $\mathcal{Z}_{0}=\mathcal{V}=1$, $\mathcal{Z}_{-}=1+q_{-}/r$, and the only correction appears in $\mathcal{Z}_{+}$ \cite{Cano:2018brq}:

\begin{equation}
\mathcal{Z}_{+}  = 1+\frac{q_{+}+\alpha' \delta q_+}{r}- \frac{\alpha' q_{+} q_-}{2 r^3 (r+q_{-})} +\mathcal{O}(\alpha'{}^{2})\, .
\end{equation}
By looking at the expression \footnote{Here $\alpha' \delta q_+<<q_+$ is an integration constant whose relation with $q_\pm$ is undetermined due to the singular behaviour of the system.}, we see that for the two-charge system $\lim_{r\rightarrow 0}\mathcal{Z}_+ \sim 1/r^3$, which is just the right behavior to obtain a horizon with non-vanishing area in \eqref{eq:metriccomp}. However, this does not fix the singularity problem, since the Kaluza-Klein scalars as well as the curvature of the full ten-dimensional solution are still divergent at $r=0$. In particular, the ten-dimensional Ricci scalar is given by 

\begin{equation}
R=\, \frac{7q_-^2}{2r^2\left(r+q_-\right)^2}\ .
\end{equation}
\noindent
Moreover, this even implies that the expression for $\mathcal{Z}_+$ cannot be trusted near $r=0$, since in its derivation it is assumed that the ten-dimensional curvature is regular at several stages. The conclusion is that the perturbative expansion in $\alpha'$ is not valid near $r=0$ when $q_{0}=q_{v}=0$, and one would need to include the full tower of higher-curvature corrections. 

From these observations we doubt there exists a \emph{true} resolution of the heterotic small black hole, as it does not seem that we can modify the structure of the fields, and a finite sized horizon built with only two active functions $\mathcal{Z_\pm}$ is headed toward a singularity in curvature. A similar analysis can also be performed for other singular solutions, like those containing three type of localized sources (say that we add S5 brane sources), with the same conclusion. Corrections of quadratic order in curvature are not sufficient to resolve the singular black hole.

\textbf{Delocalized sources in the four-charge system}:
The results in the previous section indicate that the four parameters $q_{0}, q_{+}, q_{-}, q_{v}$ must be non-vanishing if we want to describe a consistent black hole solution with a regular horizon, even if we include quadratic curvature terms. Let us therefore consider the corrections to the zeroth-order solution with $q_{0}q_{+}q_{-}q_{v}\neq 0$. As we mentioned, $\mathcal{Z}_{-}$ and $\mathcal{V}$ are uncorrected, while $\mathcal{Z}_{0}$ and $\mathcal{Z}_{+}$ read
\begin{align}
\label{eq:Z0-1}
\mathcal{Z}_{0} 
& = 
1+\frac{q_{0}}{r}- \alpha'\left[F(r;q_{0})+F(r;q_v)
\right] \, ,\\
\label{eq:Z+2}
\mathcal{Z}_{+} 
& = 
1+\frac{q_{+}}{r}\\
&+\frac{\alpha' q_{+}}{2q_v q_{0}} 
\frac{r^{2}+r(q_{0}+q_{-}+q_v)+q_v q_{0}+q_v q_{-}+q_{0}q_{-}
}{(r+q_v)(r+q_{0})(r+q_{-})}\, ,
\nonumber
\end{align}
\noindent
where  
\begin{equation}
\label{eq:Fdef}
F(r;k) 
:=\frac{(r+q_v)(r+2k)+k^{2}}{4q_v(r+q_v)(r+k)^{2}}\, .
\end{equation}
\noindent
We point out that these expressions are only valid when both $q_0$ and $q_v$ are non-vanishing.

However, we should now relate the parameters $q$ to the number of stringy objects in the system. An important property of the corrections is that they introduce delocalized sources in a way that the asymptotic charges and the near-horizon charges are different. These charges are effectively defined as the coefficient of the $1/r$ term in the functions $\mathcal{Z}_{0,\pm}$, $\mathcal{V}$ when $r\rightarrow\infty$ and when $r\rightarrow 0$, respectively. Of course, this poses the question of which of those counts the number of the corresponding stringy objects. It is particularly relevant for our discussion the case of S5 charge, codified by $\mathcal{Z}_0$. In the limits $r\rightarrow 0,\infty$, this function behaves as

\begin{equation}\label{eq:z0charge}
\lim_{r\rightarrow0}\mathcal{Z}_{0}=\frac{q_0}{r}\, ,\quad \lim_{r\rightarrow\infty}\mathcal{Z}_{0}=1+\frac{q_0-\alpha'/(2q_v)}{r}\, ,
\end{equation}

\noindent
so that near-horizon and asymptotic charges do not coincide. In the language of \cite{Marolf:2000cb}, these are respectively the ``Page charge'' and ``Maxwell charge''.

The correct way to determine the relation between those and the number of solitonic 5-branes is to couple the theory to a stack of $N$ of these branes. This can be done by dualizing the Kalb-Ramond 3-form into the NSNS 7-form $\tilde H=d\tilde B=\star e^{-2\phi} H$ and coupling the 6-form $\tilde B$ to the worldvolume action of $N$ solitonic 5-branes by means of a Wess-Zumino term, as reviewed in the supplemental material. The net effect is a localized source at the right-hand-side of the Bianchi identity (\ref{eq:bianchi}). Thus, the number of S5-branes in our solution may be computed according to
\begin{equation}
\label{eq:NS5}
N=\frac{1}{4\pi^2\alpha'}\int_{\hat{\mathbb{S}}^1\times\mathbb{S}^2_{\infty}}H-\frac{1}{16\pi^2}\int {R}_{(-)}{}^{{a}}{}_{{b}}\wedge {R}_{(-)}{}^{{b}}{}_{{a}}\, .
\end{equation}

\noindent
In the first term we used Stokes' theorem and the integral is taken on the boundary of the GH space (\ref{eq:GH}), while in the second the integral is taken over the full GH space. The result of the integration coincides with the identification in (\ref{eq:charges}) --- see the supplemental material --- and therefore it is the near-horizon charge $q_0$ the one that counts the number of S5-branes. 

On the other hand, the asymptotic charge does have a physical meaning by itself and, moreover, gives the contribution to the mass of the black hole. The negative subtraction in (\ref{eq:z0charge}) is telling us that the higher-curvature terms introduce a screening mechanism such that the charge and mass detected at infinity are smaller than the local charge produced by the 5-branes. The effective number of S5-branes detected at infinity is
\begin{equation}
\label{eq:Neff1}
N^{\rm eff}=N-\frac{2}{W}\, .
\end{equation}

\noindent
The origin of the negative shift can be identified with precision. It is caused by the presence of two self-dual gravitational instantons in the four-dimensional space parametrized by ($z$, $\vec{x}_{(3)}$), one for each $\mathfrak{so}(3) $ factor in the decomposition of the group of local Lorentz transformations $\mathfrak{so}(4) \cong \mathfrak{so}_L(3) \times \mathfrak{so}_R(3)$. The two instantons are sourced by the KK monopole and by the stack of S5 branes respectively, and each one contributes to the asymptotic charge with a factor of $-1/W$.

We obtained this result from the first-order corrected solution (\ref{eq:Z0-1}), which is expected to receive other corrections in the $\alpha'$ expansion. However, one can see that \eqref{eq:Neff1} is actually \textit{exact} and receives no further corrections \footnote{The correction associated to $\mathcal{Z}_{+}$ behaves as a delocalized source of $+2n/(NW)$ units of momentum charge, but we cannot assert that this delocalized momentum is protected under further corrections, contrary to the situation with delocalized solitonic 5-brane charge.}.

 The way to prove it is to note that the effective number of S5 branes observed at infinity is given by

\begin{equation}
\begin{aligned}
\label{eq:Neff2}
N^{\rm eff}&=\frac{1}{4\pi^2\alpha'}\int_{\hat{\mathbb{S}}^1\times\mathbb{S}^2_{\infty}}H\\
&=N+\frac{1}{16\pi^2}\int {R}_{(-)}{}^{{a}}{}_{{b}}\wedge {R}_{(-)}{}^{{b}}{}_{{a}}\, ,
\end{aligned}
\end{equation}
\noindent
where in the second line we used (\ref{eq:NS5}). But now, the integral in the second line is actually a topological invariant: it is not modified at all by continuous deformations of the connection ${\Omega}_{(-)}{}^{{a}}{}_{{b}}$, such as the ones introduced by $\alpha'$ corrections in perturbative solutions.  Hence, the value of that integral is always $-32\pi^2/W$, and the S5-brane charge measured at infinity is exactly given by (\ref{eq:Neff1}).

A very important consequence of this result is that the asymptotic S5 charge vanishes for configurations with $NW=2$.

\textbf{Fake resolution of small black holes:}
We are now ready to present a \emph{fake} resolution of the singular small black hole. Let us describe a four-charge black hole of the form (\ref{eq:H}) as a solution of the heterotic effective theory that includes all relevant terms of quadratic order in curvature.
The functions $\mathcal{Z}_{-}$ and $\mathcal{V}$ remain uncorrected as in (\ref{eq:Zs}), while $\mathcal{Z}_0$ and $\mathcal{Z}_{+}$ are given respectively by (\ref{eq:Z0-1}) and (\ref{eq:Z+2}). The solution has a regular event horizon at $r=0$, with area $A_h \propto \sqrt{nwNW}$. The near-horizon geometry is $AdS_3 \times S^3 / \mathbb{Z}_W \times \mathbb{T}^4$, and the Wald entropy is \footnote{This expression is exact in $\alpha'$ and can be obtained using the entropy function method, see \cite{Prester:2008iu}. Here $n$ should be interpreted as the total asymptotic momentum charge, whose precise relation with the momentum carried by the string is yet to be determined. Fortunately, this subtlety is not relevant for our discussion.}

\begin{equation}
\label{eq:entropy4c}
S=2\pi \sqrt{nw\left( NW+2 \right) } \, .
\end{equation}

\noindent
Here we observe a difference with the usual expression for the entropy that has appeared in the literature before --- see for instance \cite{Sen:2007qy} and references therein --- which is written in terms of the solitonic 5-brane asymptotic charge $N^{\rm eff}=N-2/W$, thus replacing the correction factor of $+2$ by a $+4$. The crucial point is that the shift between asymptotic charge and number of branes remained unnoticed, so in the preceding literature $N^{\rm eff}$ was incorrectly identified with the number of branes. There, the parameters of the near-horizon solution are identified in terms of the asymptotic charges using the zeroth-order solution. However, as we just saw, the relation between the near-horizon parameters and the asymptotic charges is altered when the higher-curvature corrections are included. It is for this reason that setting $N^{\rm eff}=0$ does not automatically imply the absence of solitonic 5-branes.

The fake small black hole is a very special four-charge solution with $NW=2$ and arbitrary $n$, $w$. It has a regular horizon and its entropy just happens to match the value \eqref{eq:microentropy}. On the other hand, it is clearly not a small black hole; it contains solitonic 5-branes and Kaluza-Klein monopole localized sources, and its asymptotic charges are different than those of \cite{Dabholkar:1989jt, Dabholkar:1990yf}. One should also notice that fake small black holes are already regular in the zeroth-order supergravity approximation.

\textbf{Discussion:}
The resolution of this system reported in \cite{Dabholkar:2004yr, Dabholkar:2004dq, Sen:2004dp, Hubeny:2004ji} considered regular near-horizon solutions in the dual frame of Type IIA on $\mathbb{K}_3 \times \mathbb{T}^2$, using a four-dimensional effective description. This phenomenon was also observed directly in the heterotic string on $\mathbb{T}^4\times\mathbb{S}^1\times \hat{\mathbb{S}}^1$ in \cite{Prester:2008iu}. In all the cases, the S5 asymptotic charge $N^{\rm eff}$ is set to zero under the assumption that this implies the absence of S5 branes. As we just saw, that premise is not true. On the other hand, in those works it is also stated that $W=0$ but, as we just argued, we found this fact to be incompatible with the assumption of a regular horizon. This incompatibility remains hidden in effective descriptions that only access partial information of the solution. This is a crucial point that, if dismissed, can produce the illusion of a stretched small horizon. Then, from all angles, it seems the solution described in these studies corresponds to the \emph{fake} small black hole we presented above. 

Our conclusions can be straightforwardly extended to five-dimensional small black holes by using ${\rm I\!R}^4$ for the Gibbons-Hawking space $d\sigma^2$. This case is simpler because there is no KK monopole and we get $N^{\rm eff}=N-1$ for the screening effect. A fake resolution of the five-dimensional small black hole is then straightforward. In this case there is just one solitonic 5-brane, whose asymptotic charge and mass vanish. The entropy is then given by $S=2\pi \sqrt{nw\left(N+2 \right)}$, which no longer happens to coincide with \eqref{eq:microentropy} for $N=1$. Notice that the resolution of five-dimensional small black holes had so far remained uncertain --- see the discussion in \cite{Sen:2009bm, Prester:2008iu}.

Notice that our results do not apply to the qualitatively different class of solutions that come into existence only after the corrections are included --- see \cite{Bueno:2012jc, Bena:2014pwa} for an explicit example. The understanding of these black holes in the context of string theory is still an open problem.

Before closing the article, we can very briefly consider the small black hole made of a Type II string carrying winding and momentum charges. We recall that in this theory the Bianchi identity does not receive corrections, so the charges are the same at the horizon and asymptotically. This means that one cannot design a fake resolution of the singularity in the terms we just described, which according to our findings clarifies why no cure for their singularity had been reported.


\textbf{Acknowledgments:}
We thank I.~Bena, S.~Chimento, R.~Emparan, O.~Lunin, S.~Mathur, R.~Russo, A.~Sen and specially T.~Ort\'in for comments and helpful discussions. 
This work was partly supported by the Italian INFN and by the Spanish grants Severo Ochoa SEV-2016-0597 and MINECO/FEDER UE FPA2015-66793-P and FPA2015-63667-P. AR is supported by ``Centro de Excelencia Internacional UAM/CSIC'' and ``Residencia de Estudiantes'' fellowships. The work of PAC is funded by Fundaci\'on la Caixa through a ``la Caixa - Severo Ochoa'' International pre-doctoral grant. PFR thanks the CERN theory division for hospitality while this work was being completed.

 
\newpage
\appendix
\onecolumngrid \vspace{1.5cm}
\begin{center}
{\Large\bf Appendices}
\end{center} \vspace{-0.4cm}
\section{The Heterotic Superstring effective action at first-order in $\alpha'$}\label{HSea}

The effective action of the Heterotic Superstring at first-order in $\alpha'$ is given by  \cite{Bergshoeff:1989de}

\begin{equation}
\label{action}
{S}
=
\frac{g_{s}^{2}}{16\pi G_{N}^{(10)}}
\int d^{10}x\sqrt{|{g}|}\, 
e^{-2{\phi}}\, 
\left\{
{R} 
-4(\partial{\phi})^{2}
+\tfrac{1}{2\cdot 3!}{H}^{2}
-\tfrac{\alpha'}{8}\left( F^A{}_{\mu\nu}F^A{}^{\mu\nu}+R_{(-)}{}_{\mu\nu}{}^a{}_b R_{(-)}{}^{\mu\nu\, b}{}_a\right)
\right\}\, .
\end{equation}
From now on we will set to zero the Yang-Mills, $F^A=0$, following the standard treatment in the small black hole literature. Here $R_{(-)}{}^a{}_b$ is the curvature of the torsionful spin connection $\Omega_{(-)}{}^a{}_b=\omega^a{}_b-\frac{1}{2}H_\mu{}^a{}_b \, dx^\mu$, namely 

\begin{equation}
R_{(-)}{}^a{}_b=d\Omega_{(-)}{}^a{}_b- \Omega_{(-)}{}^a{}_c\wedge \Omega_{(-)}{}^c{}_b\ .
\end{equation}

\noindent
The field strength $H$ of the Kalb-Ramond 2-form $B$ includes the Lorentz Chern-Simons term 
\begin{equation}
H=dB+ \frac{\alpha'}{4} \omega^{\text{L}}_{(-)}\ ,
\end{equation}
where
\begin{eqnarray}
\omega^{\text{L}}_{(-)}&=&d\Omega_{(-)}{}^a{}_b\wedge \Omega_{(-)}{}^b{}_a-\frac{2}{3}\Omega_{(-)}{}^a{}_b\wedge \Omega_{(-)}{}^b{}_c\wedge \Omega_{(-)}{}^c{}_a\ .
\end{eqnarray}
The Bianchi identity reads
\begin{equation}\label{eq:bianchi2}
dH-\frac{\alpha'}{4}R_{(-)}{}^a{}_b\wedge R_{(-)}{}^b{}_a=0\ .
\end{equation}

The $\alpha'$-corrected equations of motion are 
\begin{eqnarray}
\label{eq:eq1}
R_{\mu\nu} -2\nabla_{\mu}\partial_{\nu}\phi
+\tfrac{1}{4}{H}_{\mu\rho\sigma}{H}_{\nu}{}^{\rho\sigma}
-\frac{\alpha'}{4}R_{(-)}{}_{\mu\rho}{}^a{}_b R_{(-)}{}_\nu{}^{\rho\, b}{}_a
& = & 
0\, ,
\\
& & \nonumber \\
\label{eq:eq2}
(\partial \phi)^{2} -\tfrac{1}{2}\nabla^{2}\phi
-\tfrac{1}{4\cdot 3!}{H}^{2}
+\tfrac{\alpha'}{32}R_{(-)}{}_{\mu\nu}{}^a{}_b R_{(-)}{}^{\mu\nu\, b}{}_a
& = &
0\, ,
\\
& & \nonumber \\
\label{eq:eq3}
d\left(e^{-2\phi}\star {H}\right)
& = &
0\, .
\end{eqnarray}

\section{Wess-Zumino term for S5-branes}
\label{app:WZ}

Since solitonic 5-branes couple electrically to the dual 6-form $\tilde B$, let us start writing an action equivalent to (\ref{action}), but adapted to $\tilde B$. In a differential form language, such action reads 

\begin{equation}\label{eq:dualaction}
S=\frac{g_s^2}{16\pi G_N^{(10)}}\int \left\{e^{-2\phi}\left[\star R-4d\phi \wedge \star d\phi+ \frac{\alpha'}{4}R_{(-)}{}^a{}_b\wedge \star R_{(-)}{}^b{}_a\right]+\frac{1}{2}e^{2\phi}\tilde H\wedge \star\tilde H+\frac{\alpha'}{4} \tilde B \wedge R_{(-)}{}^a{}_b\wedge R_{(-)}{}^b{}_a \right\}\ ,
\end{equation}
where $\star e^{-2\phi} H=\tilde H\equiv d\tilde B$. 

In order to account for the coupling of solitonic five branes to the 6-form $\tilde B$, we must add to the action (\ref{eq:dualaction}) a Wess-Zumino term of the form

\begin{equation}
S_{\text{WZ}}=g_s^2 T_{S5} N \int \tilde B\ , \quad \text{where}\quad T_{S5}=\frac{1}{(2\pi \ell_s)^5 \ell_s g_s^2}\ ,  
\end{equation}
where the integral is taken over the worldvolume of the brane. This term produces a delta-like source that enters into the equation of motion of $\tilde B$ as

\begin{equation}\label{eq:eomdualB2}
d\left(\star e^{2\phi} \tilde H\right)-\frac{\alpha'}{4}R_{(-)}{}^a{}_b\wedge R_{(-)}{}^b{}_a=4\pi^2\alpha' N \star_{(4)} \delta^{(4)}(r)\ ,
\end{equation}
with $\int_{\mathcal M_4}\star_{(4)} \delta^{(4)}(r)=1$, where $\mathcal M_4$ is the four-dimensional space transverse to the S5-branes, which in the case at hands corresponds to a Gibbons-Hawking space.

We now integrate this expression, in order to obtain a relation between $q_0$ and the number of solitonic 5 branes. The first term gives 

\begin{equation}
\int_{\mathbb {\hat S}^1_\infty \times \mathbb S^2_\infty} H=\int_{\mathbb {\hat S}^1_\infty \times \mathbb S^2_\infty} \star_{(4)}d \mathcal Z_0=8\pi^2 R q_0 -\frac{8\pi^2\alpha'}{W}\ ,
\end{equation}
where $\mathbb {\hat S}^1_\infty\times \mathbb S^2_\infty$ is the boundary of the four-dimensional Gibbons-Hawking space. The computation of the second term gives the following contribution

\begin{equation}
\int_{\mathcal M_4} R_{(-)}{}^a{}_b\wedge R_{(-)}{}^b{}_a=- \int_{\mathcal M_4} d\star_{(4)}dF(r;q_v)- \int_{\mathcal M_4} d\star_{(4)}dF(r; q_0)\ ,
\end{equation}
where $F(r;k)$ is the function defined in \eqref{eq:Fdef}. The value of each term above is actually the same, since the integral is independent of the parameter $k
$, i.e.

\begin{equation}\label{eq:eomdualB}
\int_{\mathcal M_4} d\star_{(4)} dF(r;k)
=\frac{8\pi^2R_z}{q_v}=\frac{16\pi^2}{W}\ .
\end{equation}
 Then, we obtain

\begin{equation}\label{eq:instantonnumber}
\int_{\mathcal M_4} R_{(-)}{}^a{}_b\wedge R_{(-)}{}^b{}_a=-\frac{32\pi^2}{W}\ ,
\end{equation}
and, putting everything together, we get 

\begin{equation}
q_0=\frac{\alpha' N}{2 R_z}\ .
\end{equation}

 
\bibliography{biblio}{}

\end{document}